\documentclass[twocolumn,showpacs,preprintnumbers,prl]{revtex4}
\usepackage{epsfig}
\usepackage{graphicx}
\usepackage{dcolumn}
\usepackage{bm}

\begin{document}

\title{Linear-Temperature Dependence of Static Magnetic Susceptibility in
LaFeAsO from Dynamical Mean-Field Theory}
\author{S.L.~Skornyakov, A.A.~Katanin, V.I.~Anisimov}
\affiliation{Institute of Metal Physics, Russian Academy of Sciences, 620041
Yekaterinburg GSP-170, Russia \\
Ural Federal University, 620002 Yekaterinburg, Russia}
\date{\today }

\begin{abstract}
In this Letter we report the LDA+DMFT (method combining Local Density
Approximation with Dynamical Mean-Field Theory) results for magnetic
properties of parent superconductor LaFeAsO in paramagnetic phase.
Calculated uniform magnetic susceptibility shows linear dependence at
intermediate temperatures in agreement with experimental data. For high
temperatures ($>$1000 K) calculations show saturation and then
susceptibility decreases with temperature. Contributions to temperature
dependence of the uniform susceptibility are strongly orbitally dependent.
It is related to the form of the orbitally-resolved spectral functions near
the Fermi energy with strong temperature dependent narrow peaks for some of
the orbitals. Our results demonstrate that linear-temperature dependence of
static magnetic susceptibility in pnictide superconductors can be reproduced
without invoking antiferromagnetic fluctuations.
\end{abstract}

\maketitle

\textit{Introduction.} LaFeAsO \cite{Kamihara06} is the first and the most
investigated representative of novel iron pnictide superconductors. All
known to date compounds of this class: ReFeAsO$_{1-x}$F$_{x}$ (with
Re=La,Ce,Nd,Sm,Gd) and AFe$_{2}$As$_{2}$ (A=Ca,Sr,Ba) contain FeAs layers
and exhibit a common phase diagram with antiferromagnetic spin-density wave
(SDW) order appearing below $T_{\mathrm{N}}$ for parent compounds. The
antiferromagnetism is suppressed by doping or pressure with simultaneous
appearance of superconductivity. That picture resembles cuprates where
superconductivity is associated with suppression of magnetism by electron or
hole doping, although the undoped parent compounds for the
FeAs-superconductors are not Mott-Hubbard insulators but multiband metals.
These facts suggest the complex interplay of magnetism and superconductivity
in iron pnictide materials.

Recently it was found that magnetic properties of these materials show
anomalous behavior even in paramagnetic phase. A universal
linear-temperature dependence of the static magnetic susceptibility was
observed \cite{Klingeler,s2,s3,s4,s5} in both parent and superconducting
compounds in paramagnetic normal state above antiferromagnetic $T_{\mathrm{N}
}$ or superconducting $T_{\mathrm{c}}$ transition temperatures. This
non-Pauli and non-Curie-Weiss-like dependence cannot be understood within
standard theories of magnetism. The authors of \cite{Klingeler,s4} propose
that this anomalous behavior appears due to the presence of a short-range
antiferromagnetic fluctuations in the paramagnetic phase of iron pnictides.
The antifferomagnetic fluctuations were also argued to enhance the
nonanalytic (linear) term in the temperature dependence of susceptibility in
a two-dimensional Fermi liquid\cite{KorshunovPRL}, giving a possibility to
provide a contribution comparable with experimental observations.

An important issue is the effect of the Coulomb interaction on the
electronic and magnetic properties of iron pnictides \cite%
{HansmannPRL,IshidaPRB, HauleNJP}. It is generally accepted now that Coulomb
correlation effects in iron pnictides are not as strong as in cuprates where
parent compounds are wide gap Mott insulators. Experimental spectra do not
show features corresponding to Hubbard bands that are typical for strongly
correlated materials. However moderate correlations lead to a significant
renormalization of the electronic states near the Fermi energy with
effective electronic mass value $m^{\ast }/m\approx $~2 and corresponding
band narrowing of the LDA band structure is observed in ARPES \cite{ARPES}.

In this Letter we report the first \textit{ab-initio} computational results
for uniform magnetic susceptibility of parent superconductor LaFeAsO in
paramagnetic phase obtained with LDA+DMFT, which takes into account the
effect of the on-site correlations. A linear increase of magnetic
susceptibility is found below 1000 K with saturation and decreasing for
higher temperatures. As DMFT takes into account only local correlations, the
anomaly in temperature dependence of susceptibility observed in our
calculations is not connected with antiferromagnetic fluctuations. It is
shown that the observed temperature dependence of susceptibility can be
associated with narrow temperature dependent peaks in orbitally-resolved
densities of states, arising due to local correlations and located $\approx $%
100 meV below the Fermi energy. These peaks yield increase of the
susceptibility with temperature, similar to that in metamagnetic systems. In
the high-temperature regime the electrons loose their coherence and the
Curie-Weiss behavior of susceptibility is restored.

\textit{Method.}-- The LDA+DMFT scheme proceeds in two steps: (i)
construction of the effective Hamiltonian from converged LDA calculation and
(ii) solution of the corresponding DMFT equations. In the present work the
projection procedure onto Wannier functions \cite{projection} was used to
obtain an effective 22-band Hamiltonian $\hat H_{\mathrm{LDA}}(\mathbf{k})$
which incorporates five Fe \textit{d}, three O \textit{p} and three As 
\textit{p} orbitals per formula unit (two formula units per unit cell).

The DMFT self-consistent equations were solved for imaginary Matsubara
frequencies. The effective impurity problem was solved by hybridization
function expansion continuous-time quantum Monte-Carlo method (CT-QMC)\cite%
{CTQMC} with Coulomb interaction taken in density-density form. The
interaction matrix $U_{mm^{\prime }}^{\sigma \sigma^{\prime }}$ was
parametrized by parameters $U$ and $J$ according to procedure described in 
\cite{LichtAnisZaanen}. In the present work we used $U$=4~eV and $J$=1~eV
obtained via constrained LDA calculations procedure in Ref. \cite{jpcm}.
Calculations were performed in the paramagnetic state at the inverse
temperatures $\beta =$4$\div $30~eV$^{-1}$ (from 387 to 2900 K). The energy
dependence of the self-energy $\Sigma(\omega)$ on the real axis needed to
calculate spectral functions was obtained by the analytical continuation via
Pad\'{e} approximants \cite{Pade} according to procedure described in Ref. 
\cite{LaFePO10}.

Orbitally resolved DMFT densities of states were calculated as 
\begin{equation}
A_{i}(\omega )=-\frac{1}{\pi }\mathrm{Im}\sum_{\mathbf{k}}[(\omega +\mu )%
\hat{I}-\hat{H}(\mathbf{k})+\hat{\Sigma}(\omega )]_{ii}^{-1},  \nonumber
\label{SFunction}
\end{equation}%
where $\hat{H}(\mathbf{k})$ is the effective Hamiltonian and $\mu $ is the
self-consistent chemical potential. The Hamiltonian $\hat{H}(\mathbf{k})$ is
obtained by subtracting a Coulomb interaction energy $E_{dc}$ already
present in LDA (so-called double-counting correction) from \textit{d}-%
\textit{d} block of $\hat{H}_{\mathrm{LDA}}(\mathbf{k})$. The
double-counting has the form $E_{dc}=\bar{U}(n_{\mathrm{DMFT}}-\frac{1}{2})$
where $n_{\mathrm{DMFT}}$ is the total number of 3\textit{d} electrons
obtained within DMFT in the absence of the external field (such
approximation is justified for small fields used in uniform susceptibility
calculation) and $\bar{U}$ is the average Coulomb interaction parameter for
the \textit{d} shell. This choice of a double-counting term is not unique
and other variants can also be used \cite{EdcKarolak}. The form of $E_{dc}$
used in the present work yields good results for transition oxide compounds 
\cite{EdcKorotin} and pnictide materials \cite{jpcm,BaFe2As209,LaFePO10}.

The uniform magnetic susceptibility was calculated as 
\begin{equation}
\chi (T)=\frac{\partial M(T)}{\partial E_{h}}=\frac{\partial \lbrack
n_{\uparrow }(T)-n_{\downarrow }(T)]}{\partial E_{h}},  \nonumber
\end{equation}%
where $M(T)$ is the field-induced magnetization and $E_{h}$ is the energy
correction corresponding to the applied field. For all considered
temperatures the absence of polarization at zero field was checked. The
derivative was computed as a ratio of the magnetization $M(T)$ and the
energy $E_{h}$ at small fields (where $M(T)$ is a linear function of $E_{h}$%
). To minimize numerical error due to Monte-Carlo procedure for each \textit{%
T} a series of $\chi (T)$ calculations corresponding to different magnetic
fields was done.

\begin{figure}[b]
\centering \vspace{-4.0mm} \includegraphics[width=0.87\linewidth]{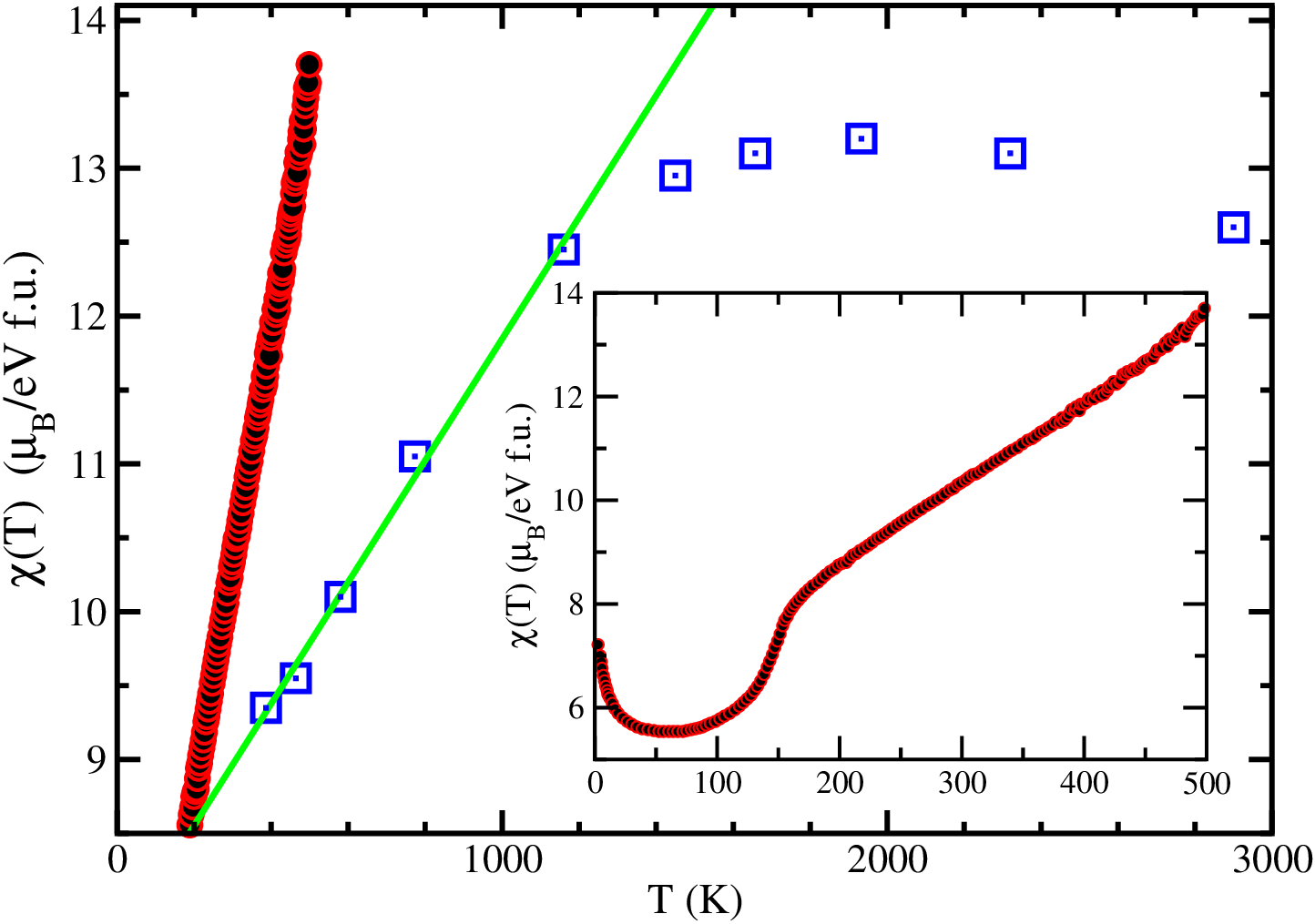}
\caption{(Color online) Uniform susceptibility $\protect\chi (T)$ of LaFeAsO
calculated within LDA+DMFT (squares) in comparison with experimental data of
Klingeler \textit{et al}.\protect\cite{Klingeler}(circles). The straight
line corresponds to least-square fit of the region below 1000 K. The inset
shows the full experimental susceptibility curve.}
\label{DMFTvsExp}
\end{figure}

\begin{figure}[b]
\centering \vspace{-4.0mm} \includegraphics[width=0.83\linewidth]{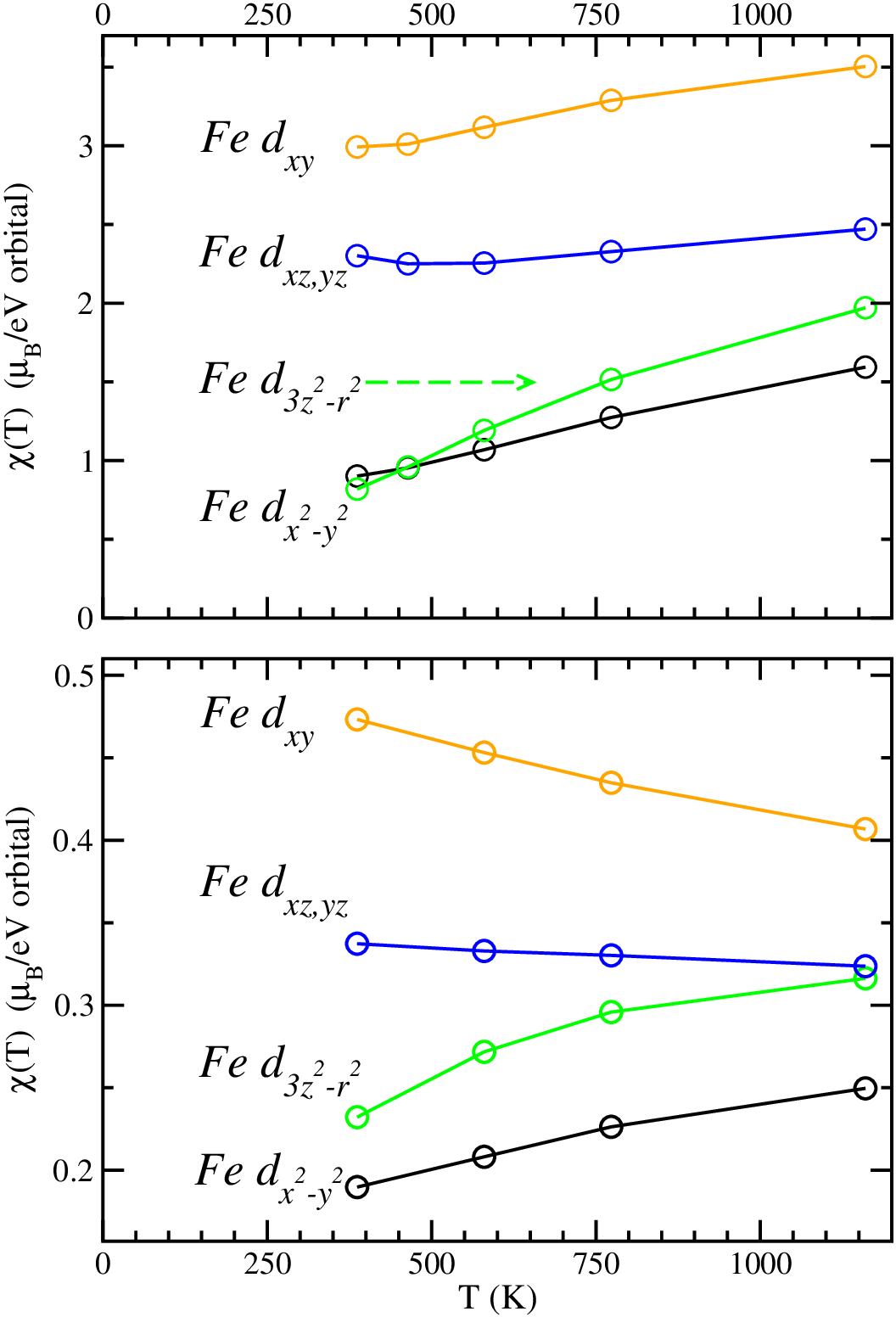}
\caption{(Color online) Orbitally resolved Fe 3\textit{d} susceptibilities $ 
\protect\chi _{m}$ vs temperature of LaFeAsO from the LDA+DMFT calculation
obtained as derivatives of the magnetization(upper panel) and estimated via
convolution of the corresponding Green's functions(lower panel).}
\label{SuscepResolv}
\end{figure}

\textit{Results and discussion.}-- In Fig.\ref{DMFTvsExp} we show calculated 
$\chi (T)$ in comparison with the experimental data of Klingeler \textit{et
al}.\cite{Klingeler}. For temperatures below $\approx $1000 K the calculated 
$\chi (T)$ demonstrates roughly linear increase. To emphasize linear
character of the susceptibility curve in low-temperature region the
least-square fit by a straight line to the obtained data is also shown in
Fig.\ref{DMFTvsExp}. For $T>1000$ K the susceptibility shows saturation and
then decreases with temperature approaching to Curie-Weiss-like behavior at
very high temperatures. To clarify the origin of the linear dependence of $
\chi (T)$ in the low temperature regime it is useful to calculate orbitally
resolved susceptibilities, $\chi _{m}=\partial M_{m}/\partial E_{h}$.
Different $d$-orbitals of iron give diverse contributions to calculated $
\chi (T)$ (see upper panel of Fig.\ref{SuscepResolv}). The contribution of $%
3z^{2}-r^{2}$ orbital posesses pronounced largest positive slope and
provides the major source for linear increase of the total susceptibility
below 1000 K. 

\begin{figure}[b]
\centering \vspace{-4.1mm} \includegraphics[width=0.9\linewidth]{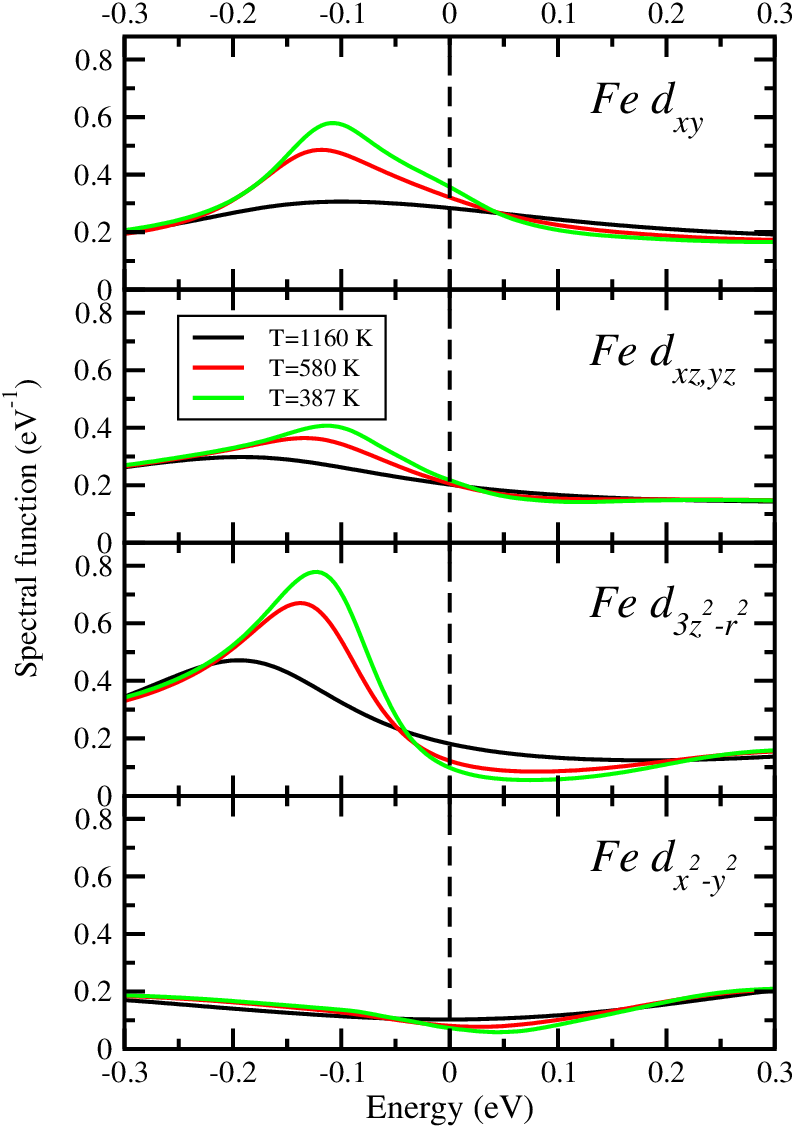}
\caption{(Color online) Temperature evolution of LaFeAsO Fe~3\textit{d}
spectral functions in vicinity of the Fermi energy (0 eV) calculated within
LDA+DMFT.}
\label{DMFTclose}
\end{figure}

It can be shown that the observed temperature dependence of $\chi _{m}(T)$
originates from pecularities of orbitally resolved densities of states $%
A_{i}(\omega ),$ obtained within LDA+DMFT (see Fig. \ref{DMFTclose}). The
calculated functions $A_{i}(\omega )$ are in good agreement with previously
published results of Aichhorn \textit{et al.}\cite{Aichhorn}. The
correlations result in renormalization of the LDA bands with corresponding
enhancement of the quasiparticle mass $m^{\ast }/m\approx $2 and lead to
appearance of peaks near the Fermi energy. The peak corresponding to $%
3z^{2}-r^{2}$ orbital is especially sharp and narrow while the peaks for
other orbitals are much more broad. The peaks $\approx $100 meV below the
Fermi energy are absent in the LDA band structure and originate purely from
the local correlations; they display strong temperature dependence,
increasing in amplitude, narrowing and approaching the Fermi level with
decreasing $T$. 

These pecularities of orbitally resolved densities of states allow to
explain qualitatively the anomalous behavior of the susceptibility. The
origin of increase of $\chi (T)$ with temperature is similar to that in
systems with van-Hove singularities near the Fermi level\cite%
{Levitin,KT,Katanin}, where increasing temperature yields activation of the
electronic states with the energy of the peak position, which results in
increase of magnetic susceptibility. When the energy $k_{\mathrm{B}}T$ ($%
T>1000$ K) becomes larger than distance from the peak to the Fermi level
(100 meV), the linear increase of uniform susceptibility stops and the curve
goes to saturation. The same mechanism applies also to the contribution of $%
x^{2}-y^{2}$ orbital, which does not show peak of the spectral function, but
a minimum very near the Fermi level and increase at larger energies.

To verify that the observed pecularities of the single-particle properties
are responsible for non-monotonic temperature dependence of $\chi (T)$, we
have estimated the uniform susceptibility as a convolution of the DMFT
interacting Green's functions neglecting vertex corrections, 
\begin{equation}
\hat{\chi}_{m}^{0}(T)=\frac{1}{\beta }\sum_{\mathbf{k,}i\omega
}\sum_{m^{\prime }}\hat{G}_{mm^{\prime }}(\mathbf{k},i\omega )\hat{G}%
_{m^{\prime }m}(\mathbf{k},i\omega ),  \nonumber
\end{equation}%
where $\hat{G}_{mm^{\prime }}(\mathbf{k},i\omega )=[(\omega +\mu )\hat{I}-%
\hat{H}(\mathbf{k})+\hat{\Sigma}(\omega )]_{mm^{\prime }}^{-1}$. Temperature
dependence of the calculated susceptibilities $\hat{\chi}_{m}^{0}(T)$ is
shown in the lower panel of Fig.\ref{SuscepResolv}. The obtained curves
qualitatively reproduce all features of the directly calculated
susceptibilities of $x^{2}-y^{2}$ and $3z^{2}-1$ orbitals, including their
slope at low temperatures (T$<$1000 K). Note that the off-diagonal
components of the Green functions also contribute partly to the linear
temperature dependence of susceptibilities. For $xy$, $xz$ and $yz$ orbitals
the vertex corrections appear to be more important.

The importance of the vertex corrections for some orbitals can be related to
the loss of the electron coherence caused by electronic correlations, which
can be traced from the electronic self-energy (Table 1). For example the
imaginary part $\Sigma _{xz,yz}$ at the Fermi level is $131$ meV at the
lowest considered temperature and increases to $306$ meV at T=$1160$ K. The
loss of the electron coherence is much less pronounced for $3z^{2}-r^{2}$
and $x^{2}-y^{2}$ orbitals where Im$\Sigma (\mathrm{E_{F}})$ are three times
smaller than the corresponding values for other orbitals at the lowest
temperature.

\begin{table}[tbp]
\caption{ Imaginary part of the self-energy (eV) at the Fermi energy E$_{%
\mathrm{F}}$ at different temperatures in LaFeAsO.}\vspace{-1.5mm} \vspace{%
3.0mm} \centering  
\begin{tabular}{c|ccccc}
\hline\hline
Temperature & 387 K & 580 K & 1160 K &  &  \\ \hline
Im$\Sigma_{xy}(\mathrm{E_{F}})$ & -0.142 & -0.242 & -0.454 &  &  \\ 
Im$\Sigma_{yz,xz}(\mathrm{E_{F}})$ & -0.131 & -0.163 & -0.306 &  &  \\ 
Im$\Sigma_{3z^2-r^2}(\mathrm{E_{F}})$ & -0.054 & -0.092 & -0.228 &  &  \\ 
Im$\Sigma_{x^2-y^2}(\mathrm{E_{F}})$ & -0.053 & -0.101 & -0.334 &  &  \\ 
\hline
\end{tabular}%
\end{table}

\begin{figure}[b]
\centering \vspace{-3.5mm} \includegraphics[width=0.87\linewidth]{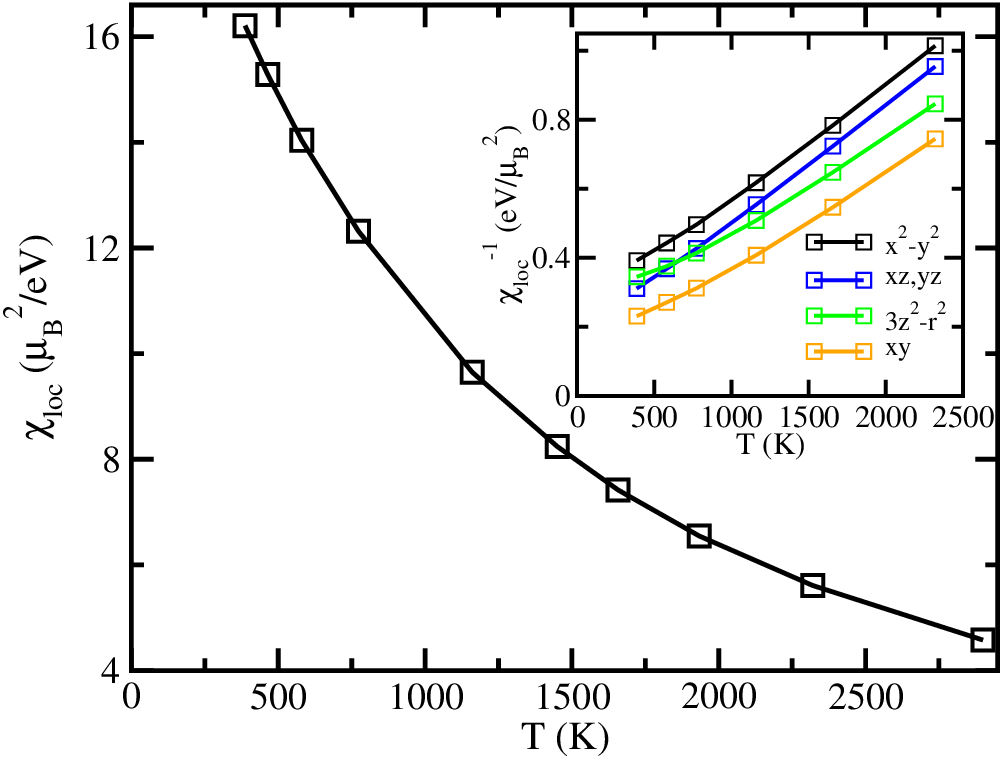}
\caption{The total local spin susceptibility versus temperature and
temperature dependence of orbitally resolved inverse of the local spin
susceptibility(inset) obtained within LDA+DMFT.}
\label{DMFTchiloc}
\end{figure}

Due to loss of coherence of electronic excitations in part of the bands, the
local susceptibility behaves very differently from the uniform
susceptibility. In Fig. \ref{DMFTchiloc} we plot temperature dependence of
the Fe-\textit{d} local spin susceptibility $\chi _{\text{loc}}(T)$ and
orbital contributions to its inverse value. The calculated $\chi _{\text{loc}
}(T)$ decreases monotonically with temperature in agreement with the results
reported in \cite{HauleNJP}. We have also calculated the mean square of the
local moment $<m_{z}^{2}>$ as a function of temperature, which is almost
temperature independent increasing approximately by only 10\% of its value
in the temperature region T=$387\div 1160$ K. Analysis of the
orbitally-resolved inverse susceptibilities (see inset of Fig. \ref%
{DMFTchiloc}) shows that the contribution of $xz$ and $yz$ orbitals is
almost linear in temperature, which goes along with loss of electron
coherence of the corresponding states, providing a possibility of the local
moment formation in these orbitals. At the same time, the electronic states
originating from $3z^{2}-r^{2}$ orbital remain more quasiparticle (Im$\Sigma
_{xz,yz}(\mathrm{E_{F}})\approx $3Im$\Sigma _{3z^{2}-r^{2}}(\mathrm{E_{F}})$%
). Such a picture is reminiscent of the proximity to the orbital-selective
Mott transition, proposed recently for iron \cite{iron}, iron pnictides\cite%
{Wu} and iron oxide\cite{iron_ox}. Quasiparticle nature of $3z^{2}-r^{2}$
states allows to justify the above discussed scenario of the linear
dependence of susceptibility due to contribution of peculiarities of the
density of states in $3z^{2}-r^{2}$ band (narrow temperature sensitive peak
in spectral functions at 100 meV below the Fermi level). 

\textit{Conclusion.}-- By employing the LDA+DMFT(CT-QMC) method we computed
paramagnetic uniform magnetic susceptibility of the parent superconductor
LaFeAsO. The obtained curve reproduces linear behavior at low temperatures
observed in experimental data of Klingeler \textit{et al.}\cite{Klingeler}.
We argue that the low-temperature linear increase of susceptibility in
LaFeAsO comes from the presence of a sharp temperature dependent peak in the
spectral function at 100 meV below the Fermi energy. Our results demonstrate
that increase of the susceptibility can be understood within single-site
dynamical mean-field approach neglecting spatial magnetic fluctuations.

\textit{Acknowledgments.}-- The authors thank J.~Kune\v{s} for providing
DMFT computer code used in our calculations and P.~Werner for the CT-QMC
impurity solver and A.V. Lukoyanov for useful discussions. This work was
supported by the Russian Foundation for Basic Research (Projects Nos.
10-02-00046a, 09-02-00431a, 10-02-00546a, and 10-02-96011ural), the Dynasty
Foundation, the fund of the President of the Russian Federation for the
support of scientific schools NSH 4711.2010.2, the Program of the Russian
Academy of Science Presidium \textquotedblleft Quantum microphysics of
condensed matter\textquotedblright\ N7, Russian Federal Agency for Science
and Innovations (Program \textquotedblleft Scientific and
Scientific-Pedagogical Trained of the Innovating Russia\textquotedblright\
for 2009-2010 years), grant No. 02.740.11.0217, the scientific program
\textquotedblleft Development of Scientific Potential of
Universities\textquotedblright\ No. 2.1.1/779.


\begin{thebibliography}{99}
\bibitem{Kamihara06} Y. Kamihara \textit{et al}., J. Am. Chem. Soc. \textbf{%
128}, 10012 (2006).

\bibitem{Klingeler} R. Klingeler \textit{et al}., Phys. Rev. B \textbf{81},
024506 (2010).

\bibitem{s2} X. F. Wang \textit{et al}., Phys. Rev. Lett. \textbf{102},
117005 (2009).

\bibitem{s3} J. Q. Yan \textit{et al}., Phys. Rev. B \textbf{78}, 024516
(2008).

\bibitem{s4} G. M. Zhang \textit{et al}., EPL \textbf{86}, 37006 (2009).

\bibitem{s5} F. Ronning \textit{et al}., J. Phys.: Condens. Matter \textbf{20%
}, 322201 (2008).

\bibitem{KorshunovPRL} M. M. Korshunov \textit{et al}., Phys. Rev. Lett. 
\textbf{102}, 236403 (2009).

\bibitem{HauleNJP} K. Haule and G. Kotliar, New Journal of Physics \textbf{11%
}, 025021 (2009).

\bibitem{HansmannPRL} P. Hansmann \textit{et al}., Phys. Rev. Lett. \textbf{%
104}, 197002 (2010).

\bibitem{IshidaPRB} H. Ishida and A. Liebsch, Phys. Rev. B \textbf{81},
054513 (2010).

\bibitem{ARPES} C. Liu \textit{et al}., Phys. Rev. Lett. \textbf{101},
177005 (2008); D. H. Lu \textit{et al}., J. Analytis \textit{et al}., Nature
(London) \textbf{455} 81, (2008); D. H. Lu \textit{et al}., Physica C 
\textbf{469}, 452 (2009).

\bibitem{projection} V. I. Anisimov \textit{et al}., Phys. Rev. B \textbf{71}%
, 125119 (2005).

\bibitem{CTQMC} P. Werner \textit{et al}., Phys. Rev. Lett. \textbf{97},
076405 (2006).

\bibitem{LichtAnisZaanen} A. I. Liechtenstein \textit{et al}., Phys. Rev. B 
\textbf{52}, R5467 (1995).

\bibitem{jpcm} V. I. Anisimov \textit{et al}., J. Phys.: Condens. Matter 
\textbf{21}, 075602 (2009).

\bibitem{Pade} H. J. Vidberg and J. W. Serene, J. Low Temp. Phys. \textbf{29}%
, 179 (1977).

\bibitem{LaFePO10} S. L. Skornyakov \textit{et al}., Phys. Rev. B \textbf{81}%
, 174522 (2010).

\bibitem{EdcKarolak} M. Karolak \textit{et al}., Journal of Electron
Spectroscopy and Related Phenomena \textbf{181}, 11 (2009).

\bibitem{EdcKorotin} Dm. Korotin \textit{et al}., Eur. Phys. J. B \textbf{65}%
, 91 (2008).

\bibitem{BaFe2As209} S. L. Skornyakov \textit{et al}., Phys. Rev. B \textbf{%
80}, 092501 (2009).

\bibitem{Aichhorn} M. Aichhorn \textit{et al}., Phys. Rev. B \textbf{80},
085101 (2009).

\bibitem{Levitin} R. Z. Levitin and A. S. Makrosyan, Sov. Phys. Uspekhi 
\textbf{31}, 730 (1988).

\bibitem{KT} S. V. Vonsovskii \textit{et al}., Fiz. Metallov. Metalloved. 
\textbf{76}, issue 3, p. 4 (1993); \textbf{76}, issue 4, p. 3 (1993).

\bibitem{Katanin} A. Katanin, Phys. Rev. B \textbf{81}, 165118 (2010).

\bibitem{iron} A. A. Katanin \textit{et al}., Phys. Rev. B \textbf{81},
045117 (2010).

\bibitem{Wu} J. Wu \textit{et al}., Phys. Rev. Lett. \textbf{102}, 126401
(2009).

\bibitem{iron_ox} A. O. Shorikov \textit{et al}., arXiv:1007.4650
(unpublished).
\end{thebibliography}
\end{document}